\documentclass[reprint,
superscriptaddress,
amsmath,amssymb,
aps,
pra,
]{revtex4-2}

\usepackage{xcolor}

\usepackage{color}
\usepackage{graphicx}
\usepackage{dcolumn}
\usepackage{bm}
\usepackage{diagbox}
\usepackage{subcaption}
\usepackage{multirow}
\usepackage{caption}
\usepackage{ragged2e}
\usepackage{float}
\usepackage{hyperref}

\DeclareUnicodeCharacter{2212}{-}

\begin{document}
\preprint{APS/123-QED}

\title{Automatic Characterization of Fluxonium Superconducting Qubits Parameters with Deep Transfer Learning}


\author{Huan-Hsuan Kung}
\email{gongsally@gmail.com}
\affiliation{Department of Physics, National Tsing Hua University, Hsinchu 30013, Taiwan}

\author{Chen-Yu Liu}
\email{d10245003@g.ntu.edu.tw}
\affiliation{Graduate Institute of Applied Physics, National Taiwan University, Taipei 10617, Taiwan}

\author{Qian-Rui Lee}
\email{ray100182181235@gmail.com}
\affiliation{Department of Physics, National Tsing Hua University, Hsinchu 30013, Taiwan}

\author{Chiang-Yuan Hu}
\email{a0935986668@gmail.com}
\affiliation{Department of Physics, National Tsing Hua University, Hsinchu 30013, Taiwan}

\author{Yu-Chi Chang}
\email{edwin4.6e13@gmail.com @gmail.com}
\affiliation{Department of Physics, National Tsing Hua University, Hsinchu 30013, Taiwan}

\author{Ching-Yeh Chen}
\email{edward035358370@gmail.com }
\affiliation{Department of Physics, National Tsing Hua University, Hsinchu 30013, Taiwan}

\author{Daw-Wei Wang}
\email{dwwang@phys.nthu.edu.tw}
\affiliation{Department of Physics, National Tsing Hua University, Hsinchu 30013, Taiwan}
\affiliation{Center for Quantum Science and Technology, National Tsing Hua University, Hsinchu 30013, Taiwan}
\affiliation{Center for Theory and Computation, National Tsing Hua University, Hsinchu 30013, Taiwan}

\author{Yen-Hsiang Lin}
\email{yhlin@phys.nthu.edu.tw}
\affiliation{Department of Physics, National Tsing Hua University, Hsinchu 30013, Taiwan}
\affiliation{Center for Quantum Science and Technology, National Tsing Hua University, Hsinchu 30013, Taiwan}
\affiliation{Taiwan Semiconductor Research Institute, Hsinchu 300091, Taiwan}

\begin{abstract}

Accurate determination of qubit parameters is critical for the successful implementation of quantum information and computation applications. In solid-state systems, the parameters of individual qubits vary across the entire system, requiring time-consuming measurements and manual fitting processes for characterization. Recent developed superconducting qubits, such as fluxonium or 0-$\pi$ qubits, offer improved fidelity operations but exhibit a more complex physical and spectral structure, complicating parameter extraction. In this work, we propose a machine learning (ML)-based methodology for the automatic and accurate characterization of fluxonium qubit parameters. Our approach utilized the energy spectrum calculated by a model Hamiltonian with various magnetic fields, as training data for the ML model. The output consists of the essential fluxonium qubit energy parameters, $E_J$, $E_C$, and $E_L$ in Hamiltonian. The ML model achieves remarkable accuracy (with an average accuracy $\sim 95.6\%$) as an initial guess, enabling the development of an automatic fitting procedure for direct application to realistic experimental data. Moreover, we demonstrate that similar accuracy can be retrieved even when the input experimental spectrum is noisy or incomplete, highlighting the model's robustness. These results suggest that our automated characterization method, based on a transfer learning approach, provides a reliable framework for future extensions to other superconducting qubits or different solid-state systems. Ultimately, we believe this methodology paves the way for the construction of large-scale quantum processors.
\end{abstract}

\maketitle

\section{Introduction}
Superconducting quantum circuits are a promising hardware platform for quantum information science, offering exceptional scalability and tunability. In recent years, multi-qubit quantum processors based on superconducting quantum circuits have been employed to demonstrate quantum supremacy and enable a variety of quantum applications\cite{Qsupremacy1, Qsupremacy2,cQED}. Superconducting qubits, the fundamental units of these circuits, typically consist of non-dissipative components including Josephson tunnel junctions, capacitors, and sometimes inductors. To effectively use superconducting qubits for quantum information tasks, precise characterization of their parameters is crucial for mitigating severe errors\cite{qubitchar1, qubitchar2}. Particularly, accurate determination of values of the energy parameters, i.e., Josephson tunnel energy $E_J$ of the Josephson junctions, charging energy $E_C$ of the capacitors, and inductive energy $E_L$ of the inductors, provide the necessary details to describe the superconducting circuits, enabling the successful execution and optimization of designed quantum operations. 

\begin{figure}
\includegraphics[scale=0.34]{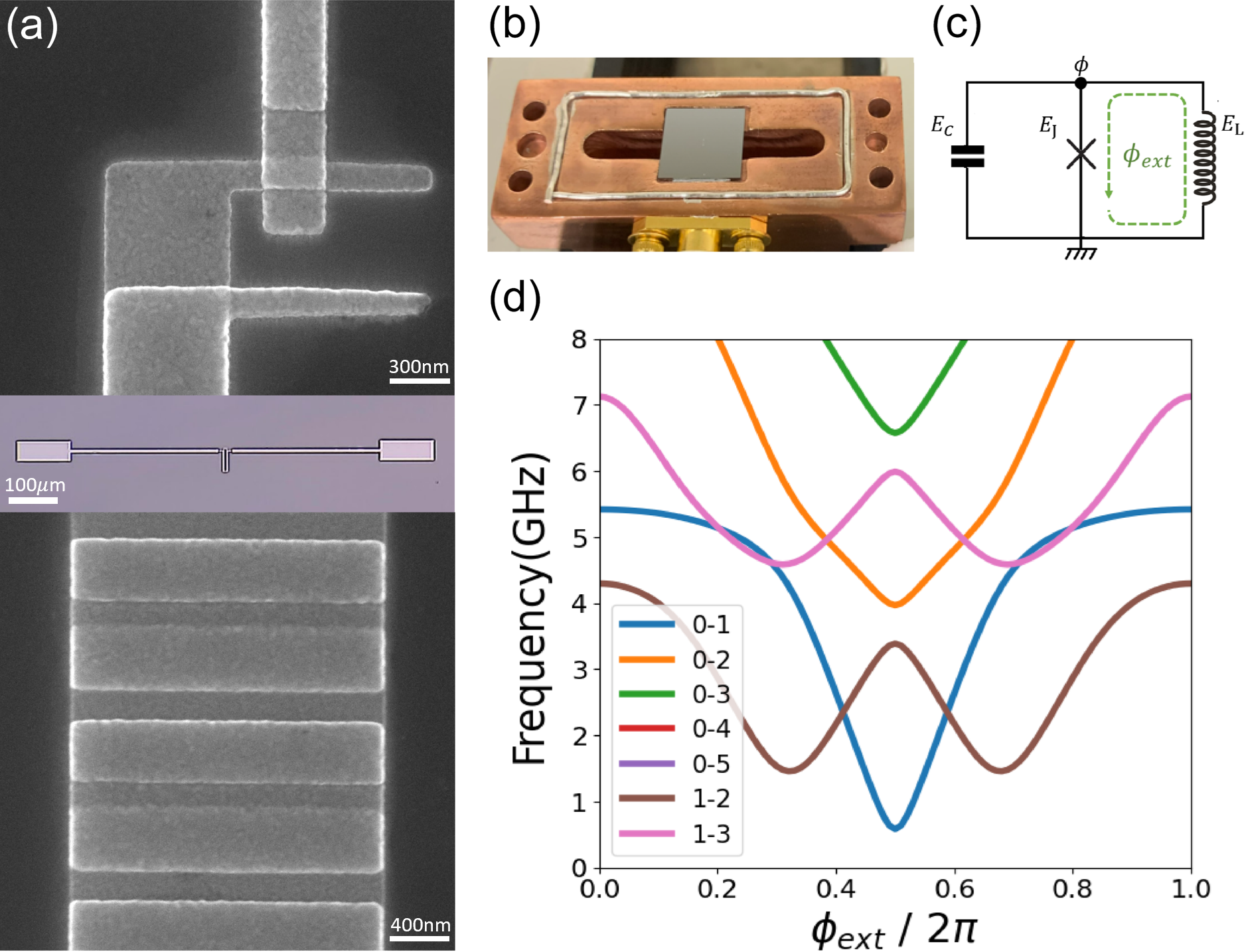}
\captionsetup{format=plain, justification=RaggedRight}
\caption{(a) Optical microscope (OM) image of a typical 3D fluxonium qubit (middle panel), along with scanning electron microscope (SEM) images of a small Josephson junction (top panel) and a segment of the Josephson junction array (bottom panel) fabricated using the Dolan bridge technique (b) A typical setup of a fluxonium qubit mounted in a three-dimensional cavity. (c) Electrical circuit model for a single fluxonium qubit, corresponding to the Hamiltonian in Eq. (\ref{eq:Hamiltonian}). (d) Theoretical energy transition spectrum, calculated from Eq. (\ref{eq:Hamiltonian}), as a function of applied external flux with energy parameters $E_J = 4.00$ GHz, $E_C = 1.00$  GHz, and $E_L = 1.00$ GHz.  
}
\label{fig: SEM image}
\end{figure}

However, due to the nonuniform fabrication of Josephson tunnel junctions, critical current variations of 5\% to 15\% across a wafer have been reported\cite{JJfabrication1}. Additionally, the Josephson tunnel barrier can age over time, leading to a drift in the magnitude of the critical current. Thus, $E_J$ may deviate by as much as 30\% from its designed value\cite{JJfabrication2}. This makes the actual energy parameters of the fabricated qubits uncertain, necessitating precise characterization. A typical method to extrapolate the associated energy parameters is to measure the energy transition spectrum of each qubit and then fit to the theoretical model Hamiltonian using multiple fitting parameters. For large-scale multi-qubit quantum processors, efficiently and precisely characterizing these energy parameters for every single qubit becomes a complex challenge. Hence a fully automatic characterization process is essential. Moreover, several recently developed superconducting qubits, for example, fluxonium qubits\cite{fluxonium1} (see Fig. \ref{fig: SEM image}) or 0-$\pi$ qubits, have been designed to exploit intrinsic noise protection but feature more complex transition spectra\cite{Noise_protected}. These noise-protected devices have the potential to be scaled up for multiqubit quantum processors\cite{fluxonium2}, but the automatic characterization process is even more demanding. Current existing tools for parameter characterization of complex superconducting qubits, such as QFit within the scQubit package \cite{scQubit_Qfit}, still require many time-consuming manual steps. As a result, the capability of automatic and efficient characterization for multi-qubit systems is still limited in the present development.

In this work, we introduce a machine learning (ML)-augmented fitting approach to improving the efficiency of characterizing the energy parameters of a single superconducting fluxonium qubit. The methodology leverages ML to assist in the initial estimation and transition identification, facilitating an automated fitting process to determine the energy parameters, $E_J$, $E_C$, and $E_L$ of the single fluxonium qubit's Hamiltonian. By using only the calculated qubits transition spectra from the model Hamiltonian for the training process, our approach achieves an initial guess for the fluxonium qubit energy parameters with an average accuracy $\sim 95.6\%$. This significantly accelerates the convergence of the fitting process when applied to realistic experimental data. Furthermore, our approach can successfully characterize the energy parameters even using partial or fuzzy spectra, greatly reducing the measurement time required for qubit parameter characterization. Consequently, our work demonstrates the successful automation of parameter identification for complex superconducting qubits, eliminating the need for manual parameter searches and speeding up the development of a large-scale multi-qubit quantum processor. 

The structure of this paper is organized as follows: In Sec. \ref{sec:Qubit Parameter Identification}, we provide an overview of fluxonium qubit parameter characterization. Sec. \ref{sec: Data Preparation} outlines the preparation of the training and test data, while Sec. \ref{sec: Model Design and Training} presents a detailed description of machine learning model design and training. In Sec. \ref{sec:Results}, we highlight several our key results: Sec. \ref{subsec:The prediction from the machine learning model} and Sec. \ref{subsec: Error and Cost Distributions of Traditional Fitting Methods} cover the prediction performance of the machine learning model. Sec. \ref{subsec:Comparison between ML Model and Traditional Method}  compares these results with traditional fitting methods. Finally, Sec. \ref{subsec:The application on real fluxonium qubit energy spectrum} demonstrates the application of our ML approach in an automatic process on a real fluxonium qubit energy spectrum. After a detailed discussion in Sec. \ref{sec:Discussion}, we conclude our work in Sec. \ref{sec:Conclusion}. 

\section{Parameters Characterization of Fluxonium Qubits: Standard Method and New Approach}
\label{sec:Qubit Parameter Identification}
In this section, we first describe the standard method for the parameter characterization of a fluxonium qubit in Sec. \ref{subsec: Standard Parameter Characterization Method}. We then provide an overview of our new approach to automatic characterization using deep transfer learning in Sec. \ref{subsec:An overview of our Method}. Subsequently, we describe the details of the latter method and compare the obtained results in the following sections.
\subsection{Standard Parameter Characterization Method}
\label{subsec: Standard Parameter Characterization Method}

A typical method for characterizing superconducting qubit parameters involves fitting the measured energy transition spectrum to the transition energies derived from the modeling Hamiltonian of the qubit. Using a single fluxonium qubit as an example, the Hamiltonian consists of a single Josephson junction shunted with a capacitor and a super-inductor:
\begin{eqnarray}
    \hat{H} = 4E_C \hat{n}^2 - E_J \cos (\hat{\phi} + \phi_{\text{ext}}) + \frac{1}{2} E_L \hat{\phi}^2,
\label{eq:Hamiltonian}
\end{eqnarray}
where $E_C =e^2/2C$ is the charging energy (with $C$ being the capacitance), and $E_L = \left(\frac{\hbar}{2e}\right)^2 \frac{1}{L}$ is the inductive energy (with $L$ being the inductance). Here, $\hat{\phi}$ is the phase twist operator across the inductance, and $\hat{n}$ is the displacement charge operator. $\phi_{\text{ext}}$ is the reduced magnetic flux biasing the loop formed by the weak junction and the shunting inductance \cite{fluxonium0}. Figure \ref{fig: SEM image} shows an experimental setup of a single fluxonium in a 3D cavity, along with the corresponding transition energy spectrum.

\begin{figure*}
    \centering
        \centering
        \includegraphics[scale=0.45]{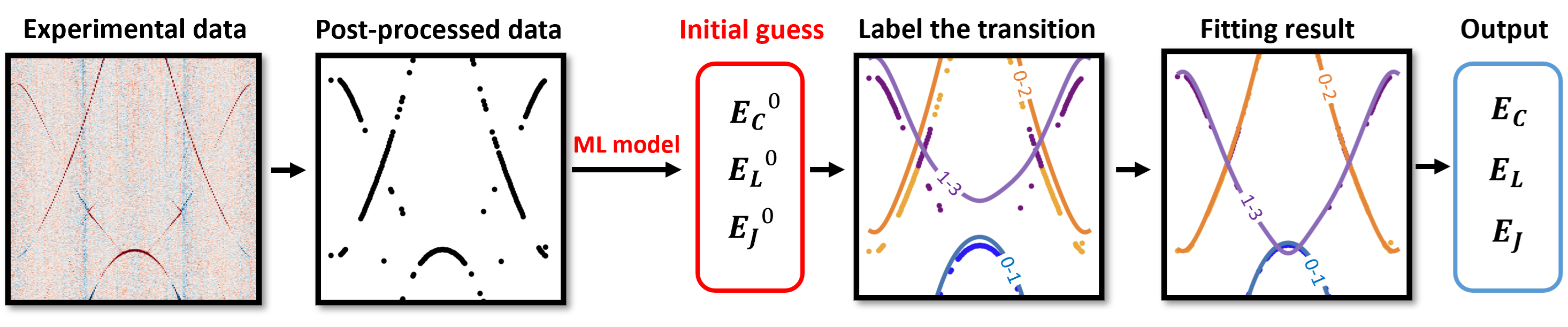}
        \captionsetup{format=plain,justification=RaggedRight}
        \caption{Flowchart of the qubit parameter identification method proposed in this work. The experimental data is firstly post-processed and then fed into the ML model to generate the initial guess of the qubit parameters, \(E_C^0\), \(E_L^0\), and \(E_J^0\). Based on these initial parameters, the post-processed data can be automatically labeled as various spectrum transition lines, which can be utilized in the final rule-based fitting process (see the text for more details). Note that the initial guess of \(E_C^0\), \(E_L^0\), and \(E_J^0\), can also be used as the starting point in the final fitting process to accelerate the fitting process and to improve the final accuracy.}
        \label{fig:flow}
\end{figure*}

To characterize the key energy parameters \(E_C\), \(E_L\), and \(E_J\) from experimental data, a transition spectrum is typically measured using a dispersive readout scheme over a few GHz, as a function of the external flux across a single flux quantum period. To simplify the fitting process and modeling, the spectrum is measured away (at least 100 MHz in our case) from the resonance frequency of the readout resonator, considering only the Hamiltonian of a single fluxonium qubit. Data points corresponding to a specific transition are then selected, and a least-squares fitting method is applied to match the simulated results from the fluxonium Hamiltonian. However, the identification of transitions and data selection are typically performed manually. Furthermore, the use of a random initial guess in this numerical method can result in slow convergence rates or inaccurate circuit parameters. This process requires significant manual effort and time to identify a proper initial guess, posing challenges for parameter determination. 

\subsection{An overview of our Automatic Parameter Characterization Method}
\label{subsec:An overview of our Method}
Here we present an overview of our novel approach that utilizes a deep transfer learning scheme to automate and accelerate the characterization of a single fluxonium qubit's energy parameters, $E_C$, $E_L$, and $E_J$. By leveraging the power of ML algorithms, our goal is to automatically provide an initial estimation process and transition spectrum identification, ultimately improving the accuracy and efficiency of extracting crucial qubit parameters.

The procedure, as outlined in Fig.~\ref{fig:flow}, begins with the experimentally measured transition energy spectrum, which typically includes several distinct transitions as a function of external flux. This data undergoes post-processing that includes noise filtering and extrapolation of transition spectrum points. To enhance the accuracy of extrapolating transition points, a band-pass filter is first applied to data points according to the measured signal magnitude. A typical selection window is higher than two and a half standard deviations of the background average and lower than 20\% of the maximum measured magnitude. Next, the Python package find\_peaks\_cwt  is employed to pinpoint data points of the transition spectrum at magnitude extrema. These processed points are fed into a pre-trained ML model, which will be described in detail later, to generate initial guesses for \(E_C^0\), \(E_L^0\), and \(E_J^0\). A simulated spectrum using these initial guesses is employed to automate the identification of the transitions of processed data. The processed points are then labeled as belonging to a transition only if a single simulated transition is within 0.3 GHz. Points that are either far away from any simulated transition or fall within a region containing multiple simulated transitions within 0.3 GHz are excluded as outliers. In the final stage, the labeled points are subjected to a least-squares fitting routine. The values of \(E_C^0\), \(E_L^0\), and \(E_J^0\) serve as the initial guesses for this process, with the single fluxonium Hamiltonian to obtain the best-fitting parameters  \(E_C\), \(E_L\), and \(E_J\). This systematic approach ensures both precision and efficiency of the transition identification and automated characterization processes.

\begin{figure*}
        \centering
        \includegraphics[scale=0.45]{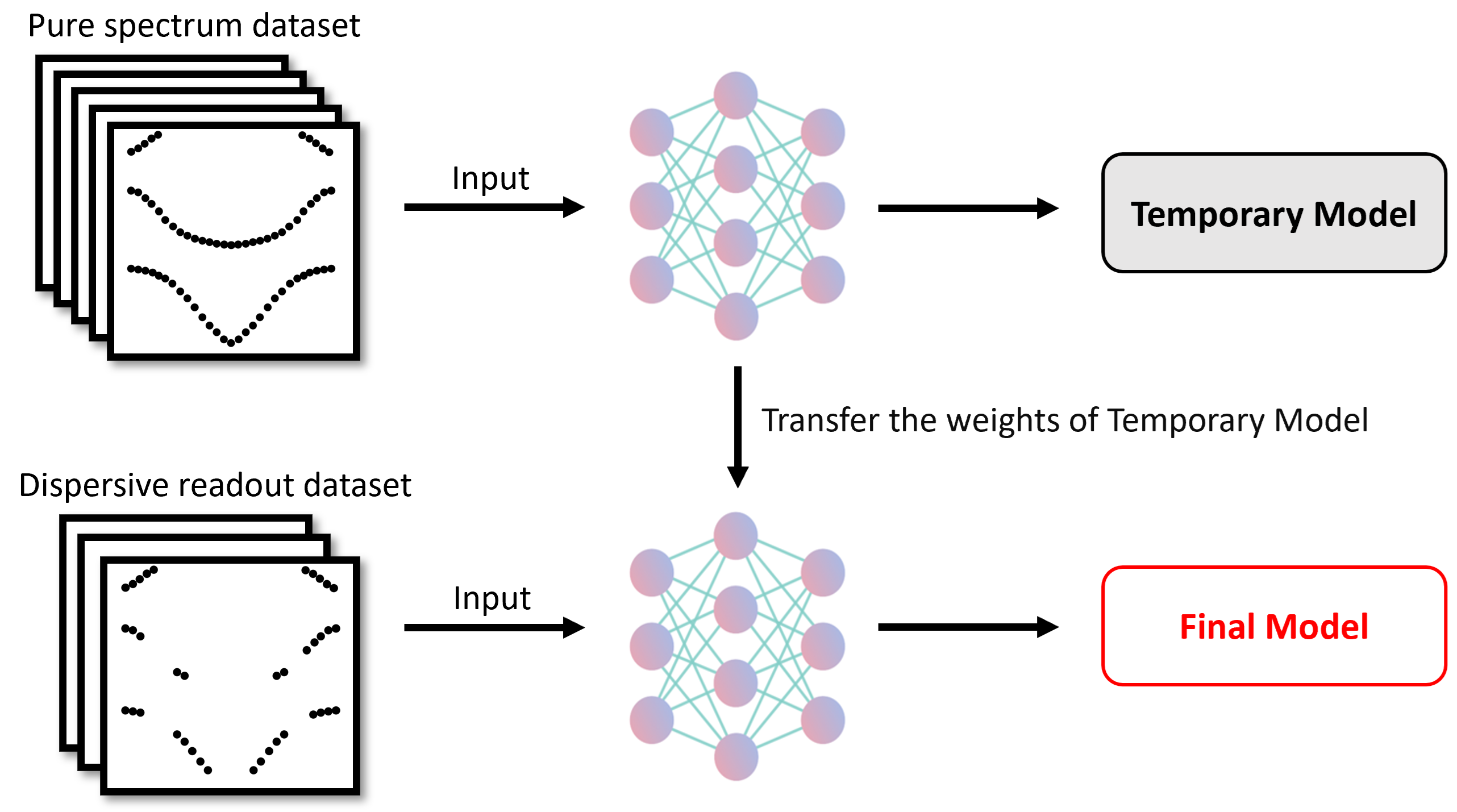}
        \captionsetup{format=plain,justification=RaggedRight}
        \caption{The flowchart of our ML model building process. Initially, a large amount of pure spectrum data is employed to pre-train a deep neural network model, which captures the fundamental characteristics of the data in the model parameters. The second stage is to undergo a fine-tuning process, where all the model parameters are fixed as the pre-trained result except for the parameters of the last level of neurons. These parameters are then trained using the dispersive readout dataset (see the text). This procedure allows the Final Model to inherit broad knowledge from the pre-trained model but provides a more accurate output when compared to the realistic experimental measurements.}
        \label{fig:flow_AI}
    \label{fig:flow_ML}
\end{figure*}
\section{Preparation of Training Data}
\label{sec: Data Preparation}

To provide an accurate and efficient initial estimation of circuit parameters, we propose a specialized ML approach that utilizes a deep transfer learning model, trained on simulated single fluxonium transition spectra, as illustrated in Fig.~\ref{fig:flow_AI}. We define two types of simulated training datasets: the pure spectrum dataset and the dispersive readout dataset, described in the following paragraphs. All simulated data presented in this work are generated within the frequency range of $4.0-8.0$ GHz, with the energy transitions considered being $0-1$, $0-2$, $0-3$, $0-4$, $0-5$, $1-2$, and $1-3$ (see Fig.~\ref{fig: SEM image}(d)). The readout resonator frequency is set to $6.0$ GHz, which is the typical design value in our experimental setup. The qubit parameter ranges for all the training datasets are \(E_C \in [0.5, 3.0]\) GHz, \(E_L \in [0.1, 2.0]\) GHz, and \(E_J \in [2.0, 10.0]\) GHz, selected to reflect the commonly used operational conditions of fluxonium qubits and corresponding fabrication parameters, as reported in the literature \cite{fluxonium0,fluxonium1,fluxonium2}.

The pure spectrum dataset is prepared for training our ML model with different combinations of  \(E_C\), \(E_L\), and \(E_J\). For each combination, the spectrum is calculated using the Quantum Toolbox in Python (QuTip), based on the transition energies between different energy levels according to the Hamiltonian in Eq.~\ref{eq:Hamiltonian}, while varying the external phase $\phi_{\text{ext}}$ with a resolution of 256 points per flux period. The resulting spectrum at a given external flux is plotted as ``black dots" for each transition, as shown as the ``Pure spectrum dataset" of Fig.~\ref{fig:flow_AI}. This pure spectrum dataset does not include any coupling terms, allowing for the generation of a large amount of training data in a short time. Furthermore, these datasets show good agreement with real single-qubit spectra when measured away from the frequency regime coupling with readout resonator or other qubits. In total, we used 15392 pure spectrum datasets to train our model.

The dispersive readout dataset, which includes the dispersive readout mechanism, is designed for the second stage of our deep transfer learning process, requiring higher precision simulations. The magnitude of the readout voltage shift for each transition, due to the dispersive shift associated with coupling to the readout resonator, is calculated using second-order perturbation theory. Specifically, we consider a Lorentzian response of a $6.00$ GHz readout resonator with a line width of $7$ MHz and the coupling strength is $g=100$ MHz. We then compute the voltage change in the readout response caused by the dispersive shift for a saturation drive at every transition and flux value. Data points where the readout voltage change is less than 10\% of the magnitude at readout resonance are excluded. The resulting spectrum is shown as the ``Dispersive readout dataset" in Fig.~\ref{fig:flow_AI}. Due to the complexity of the computation, the time required to generate a dispersive readout dataset is typically more than 100 times longer than for the pure spectrum dataset for a single qubit parameter, resulting in a significantly smaller quantity compared to the simple dataset. To complete the Final Model, we use the deep transfer learning process with 469 dispersive readout datasets. 

\section{Model Design and Training}
\label{sec: Model Design and Training}

In the traditional fitting approach, the goal is to determine the values of \(E_C\), \(E_L\), and \(E_J\) that accurately replicate a given energy transition spectrum. From a ML perspective, this fitting task is equivalent to identifying the functional correlation between the transition spectrum dataset, represented by \(\mathbf{S}_E\), and the energy parameters of the Hamiltonian (see Eq. \ref{eq:Hamiltonian}). Here, the objective is to determine a function \(F\) such that \(F(\mathbf{S}_{E})=\mathbf{E}\), where $\mathbf{E}\equiv\left[ E_C, E_L, E_J \right]$ denotes the vector of these parameters. To establish this complex functional relationship \(F\), a neural network model \(F_{\text{NN}}\) can be employed. This model is trained to approximate \(F\) effectively, facilitating an initial guess of energy parameters directly from the spectrum data:
\begin{eqnarray}
F_\text{NN} \Rightarrow F(\mathbf{S}_{E}) = \mathbf{E}.
\end{eqnarray}

The loss function is designed as the mean square error (MSE) of the model predictions and the actual values of $E_C$, $E_L$, and $E_J$ in the training data:
\begin{eqnarray}
    \text{Loss} = \frac{1}{N_{\text{train}}} \sum_{i = 1}^{N_{\text{train}}}\left(F_{\text{NN}}(\mathbf{S}_{E}^i) - \mathbf{E}^i\right)^2,
\end{eqnarray}
where the superscript $i$ indicates the different combinations of $E_C$, $E_L$, and $E_J$ in the training data. According to the Universal Approximation Theorem \cite{uat1}, it is reasonable to believe that the discrepancy between the approximated function ($F_\text{NN}$) and the true function ($F$) can become negligible if the number of fitting parameters and the training data used for $F_\text{NN}$ are sufficiently large. Even with a smaller amount of training data $F_\text{NN}$ can still provide a close estimation of the values of  $E_C$, $E_L$, and $E_J$, which can then be used as an initial guess for the detail fitting process. 

The Swin Transformer V2 neural network model \cite{swinV2} was selected to implement $F_\text{NN}$, simulating the functional relationship $F$. The model parameters were optimized using the Prodigy optimizer \cite{mishchenko2023prodigy}. Compared to other popular machine learning models, such as the ResNet \cite{resnet1} and DenseNet \cite{densenet1} series, Swin Transformer V2 is relatively lightweight. This smaller model size could be advantageous for future deployment on compact hardware near qubits, facilitating efficient integration with quantum computing systems.

\begin{figure*}
    \includegraphics[scale=0.38]{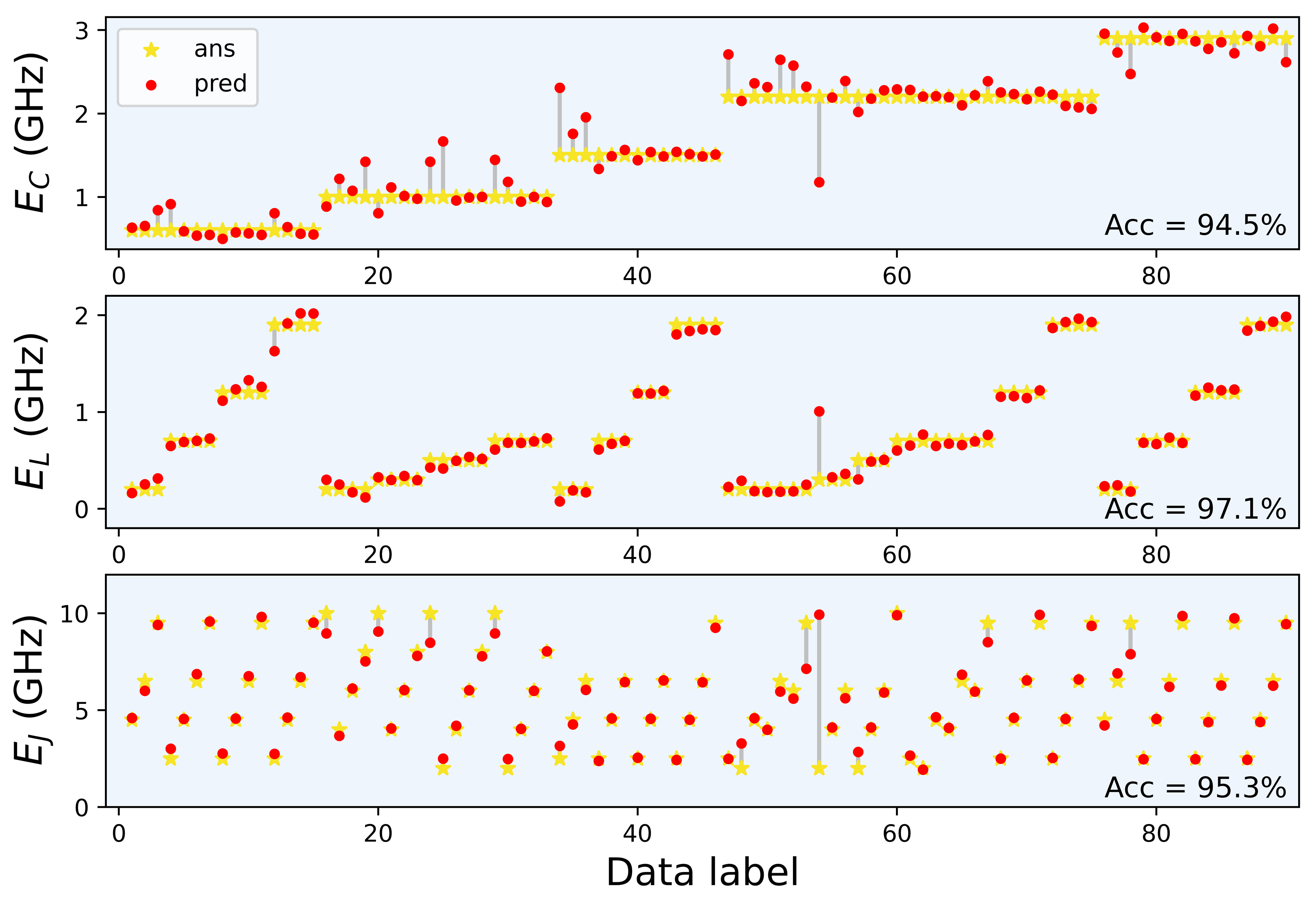}
    \captionsetup{format=plain,justification=RaggedRight}
    \caption{
    The predictive performance of our ML model for the fluxonium qubit energy parameters: (a) charging energy $E_C$, (b) inductive energy $E_L$, and (c) Josephson energy $E_J$. The model was evaluated on a test dataset comprising $N_{\text{test}} = 512$ samples, of which a representative subset of 90 data points is presented for visual clarity.  
    }
    \label{fig:pred_res}
\end{figure*}

Simulating an energy transition spectrum that closely matches real measured data is typically time-intensive due to the detailed simulations required, such as the dispersive shift of each transition. However, a simpler spectrum that retains only transition energies, omitting details of spectral weights, can be generated more efficiently. To facilitate this, we have implemented a two-step deep transfer learning methodology \cite{tf1}, which effectively utilizes both pure spectrum datasets ($N_{\text{train}}^{p} = 15392$) and dispersive readout datasets ($N_{\text{train}}^{d} = 469$). This TL approach leverages the Prodigy optimizer \cite{mishchenko2023prodigy}, a parameter-free learning algorithm, and enhances the Swin Transformer V2's capability to discern critical features necessary for determining qubit parameters from the experimental energy transition spectrum. This method proves particularly effective when the available detailed simulation data is limited.

\section{Results}
\label{sec:Results}

\subsection{Prediction results on Qubit Parameters}
\label{subsec:The prediction from the machine learning model}

From the setup described previously, we now present the prediction results obtained using our ML model. For testing, we generated test a dataset of 512 spectra derived from non-repetitive combinations of \(E_C\), \(E_L\), and \(E_J\), distinct from those used during training. These parameter values lie within the typical experimental range observed for fluxonium qubits, consistent with the training dataset described previously. The goal is to test whether our ML model can provide accurate initial guesses of energy parameters based on their corresponding energy spectra.

To quantitatively evaluate the accuracy of the ML model's predictions for $E_C$, $E_L$, and $E_J$, we define the average accuracy as follows
\begin{eqnarray}
    Acc(E_\nu) \equiv  \frac{1}{N_{\text{test}}}\sum_{i = 1}^{N_{\text{test}}} \left( 1 -  \frac{| E_{\nu}^{i} - E_{\nu}^{\text{true},i}|}{\text{R}(E_{\nu}^{\text{test}})}  \right),
    \label{Eq:Accuracy_def}
\end{eqnarray}
where $E_{\nu}^{i}$ is the predicted value from the ML model and $E_{\nu}^{\text{true},i}$ is the true energy parameter for the $i$th test data with $\nu$ standing for $C$, $L$, or $J$ respectively. The term $R({E}_{\nu}^{\text{test}})$ denotes the parameter range used in the training dataset, which is 2.5 GHz for $E_C$, 1.9 GHz for $E_L$, and 8.0 GHz for $E_J$ in our case. Notably, this definition of accuracy in Eq. (\ref{Eq:Accuracy_def}) differs from either the traditional definition for discrete labels (i.e. classification) or the mean absolute error (MAE) for continuous labels. By normalizing the deviation relative to the test data range, our definition provides a more practical basis for comparison between different ranges of energy parameters under current experimental conditions. For example, in our case, a 95\% accuracy represents an average deviation of 5\% in the parameter range, and hence indicates the frequency deviation of 0.125 GHz, 0.095 GHz, and 0.4 GHz for $E_C$, $E_L$, and $E_J$ respectively.

Figure ~\ref{fig:pred_res}(a)-(c) present the representative predicted values (red dots) of $E_C$, $E_L$, and $E_J$ respectively compared with the true values of energy parameters from the test dataset (yellow stars). To improve visual clarity, we display only a representative subset of 90 datasets out of the total 512 test cases.  Our model achieves high average prediction accuracies of 94.5\% for \(E_C\), 97.1\% for \(E_L\), and 95.3\% for \(E_J\), yielding an overall averaged accuracies of 95.6\% across all three energy parameters. It is worth noting that the test cases with accuracy below 90.0\% typically lack sufficient transition spectrum points within the $4.0-8.0$ GHz range. Additionally, we have to emphasize that the predicted value, \(E_C^0\), \(E_L^0\), and \(E_J^0\) from ML model are not the final qubit energy parameters. However, their high accuracy serves two critical purposes: first, they provide reliable initial guesses for subsequent fitting steps, preventing convergence to local minima that cause significantly deviated qubit parameters. Second, they enable an automated approach for spectrum identification, streamlining the parameter characterization process.    

\subsection{Error and Cost Distributions of Traditional Fitting Methods}\label{subsec: Error and Cost Distributions of Traditional Fitting Methods}

To quantitatively assess the efficacy of using an initial guess from our ML approach in comparison to a random guess, we conducted a comparative analysis of the subsequent fitting results and cost function using the test dataset of spectra with known qubits' parameters. The fitting process was constrained to five iterations, utilizing a combination of initial guesses in the range of ML model training dataset. To evaluate the overall error compared to the true parameters, we define a single value error function that equally contributed by the accuracy of $E_C$, $E_L$, and $E_J$:  
\begin{eqnarray}
    Error \equiv 1-\frac{1}{3}\sum_{\nu=C,L,J}Acc(E_\nu). 
\end{eqnarray}
We further introduced another single value metric, $Cost$ function, to measure the averaged discrepancy between the fitted spectrum and the correct energy transition spectrum:
\begin{eqnarray}
    Cost \equiv \frac{1}{N} \sum_{i = 1}^{N} \left( f(\phi_i)-f_i \right)^2,
\end{eqnarray}
where $N$ is the total number of data points in the energy transition spectrum of all $0-1$, $0-2$, $0-3$, $0-4$, $0-5$, $1-2$, and $1-3$ transitions (illustrated in Fig. \ref{fig: SEM image}(d)) for a given set of parameters $(E_C,E_J,E_L)$. The variables $\phi_i$ and $f_i$ denote the external phase ($\phi_{\text{ext}}$) and transition frequency, respectively, for the $i$th data point in a spectrum. The function $f(\phi_i)$ represents the transition frequency calculated by solving the system Hamiltonian in Eq. (\ref{eq:Hamiltonian}) using a given set of $(E_C,E_J,E_L)$, which may be derived from either our ML model approach or traditional fitting methods. For illustration, we choose a spectrum of a qubit with parameter $(E_C,E_J,E_L)=(1.28 \text{ GHz}, 1.50\text{ GHz}, 0.70\text{ GHz})$ as an example case.

\begin{figure}[h]
\begin{subfigure}{0.23\textwidth}
\includegraphics[scale=0.26]{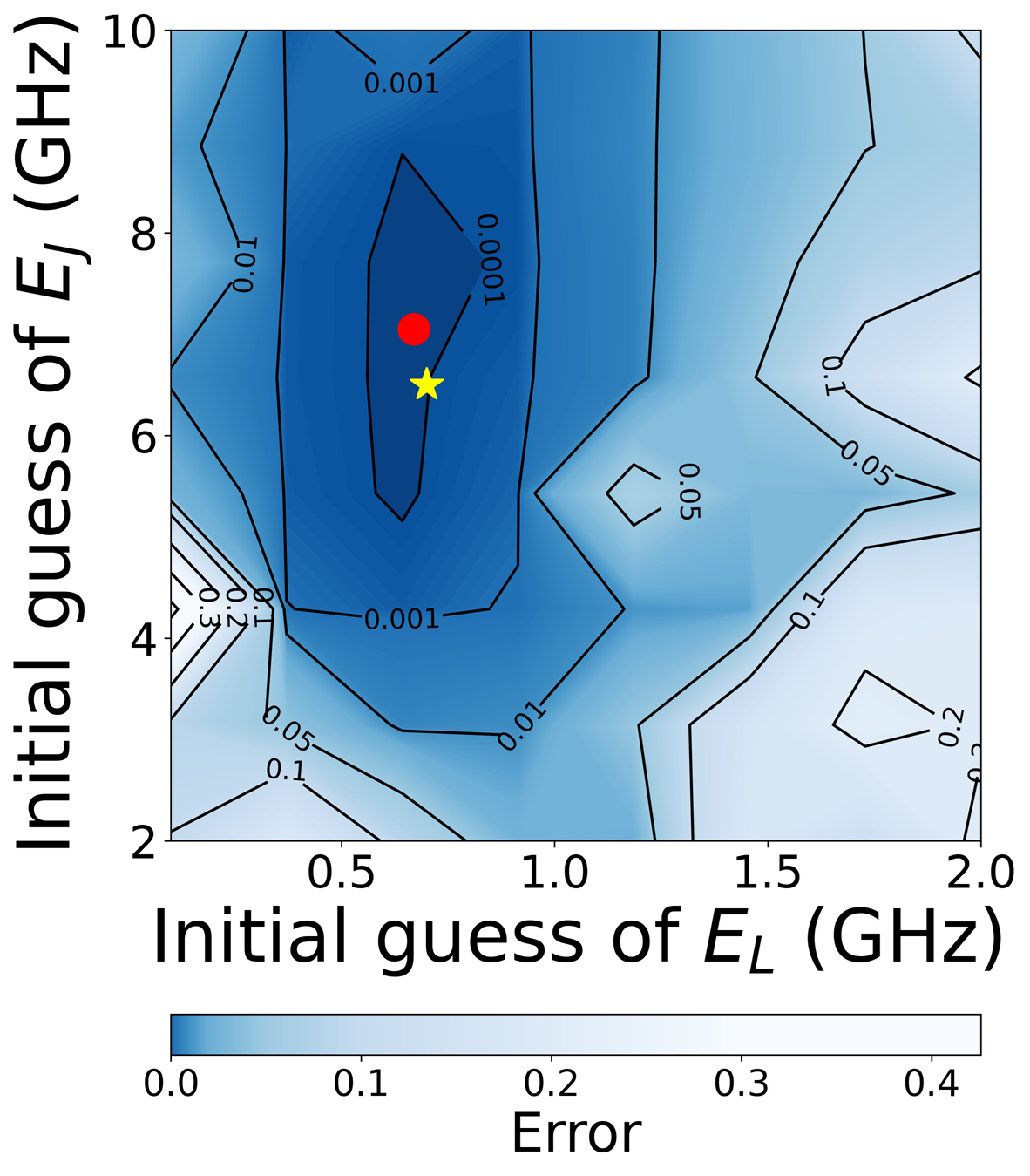} 
\caption{$Error$}
\label{fig:difference_ElEj}
\end{subfigure}
\begin{subfigure}{0.23\textwidth}
\includegraphics[scale=0.26]{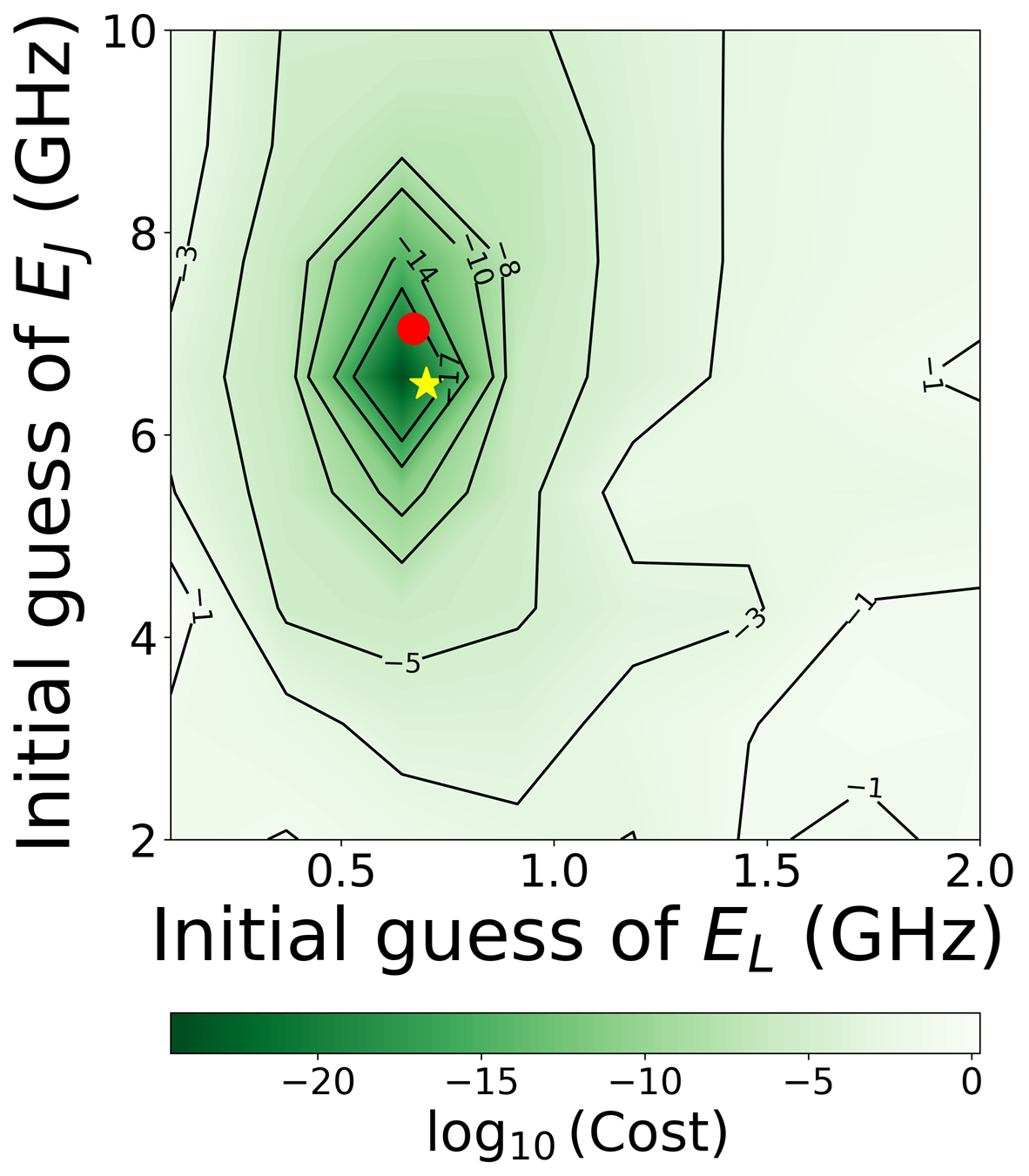}
\caption{$\log_{10}(Cost)$}
\label{fig:cost_ElEj}
\end{subfigure}
\captionsetup{format=plain, justification=RaggedRight}
\caption{Contour plots of $Error$ and $Cost$ obtained by transitional fitting methods with 5 iterations for various initial values in the $E_L-E_J$ space. Here $E_C=1.28$ GHz is fixed for the convenience of representation. The correct qubit parameters are $E_C=1.5$ GHz, $E_L=0.7$ GHz, $E_J=6.5$ GHz as labeled by the yellow star symbol. The red dot at $E_L=0.67$ GHz and $E_J=7.05$ GHz represents the results provided by the ML model.}
\label{fig:ElEj}
\end{figure}

\begin{figure}[h]
\begin{subfigure}{0.23\textwidth}
\includegraphics[scale=0.26]{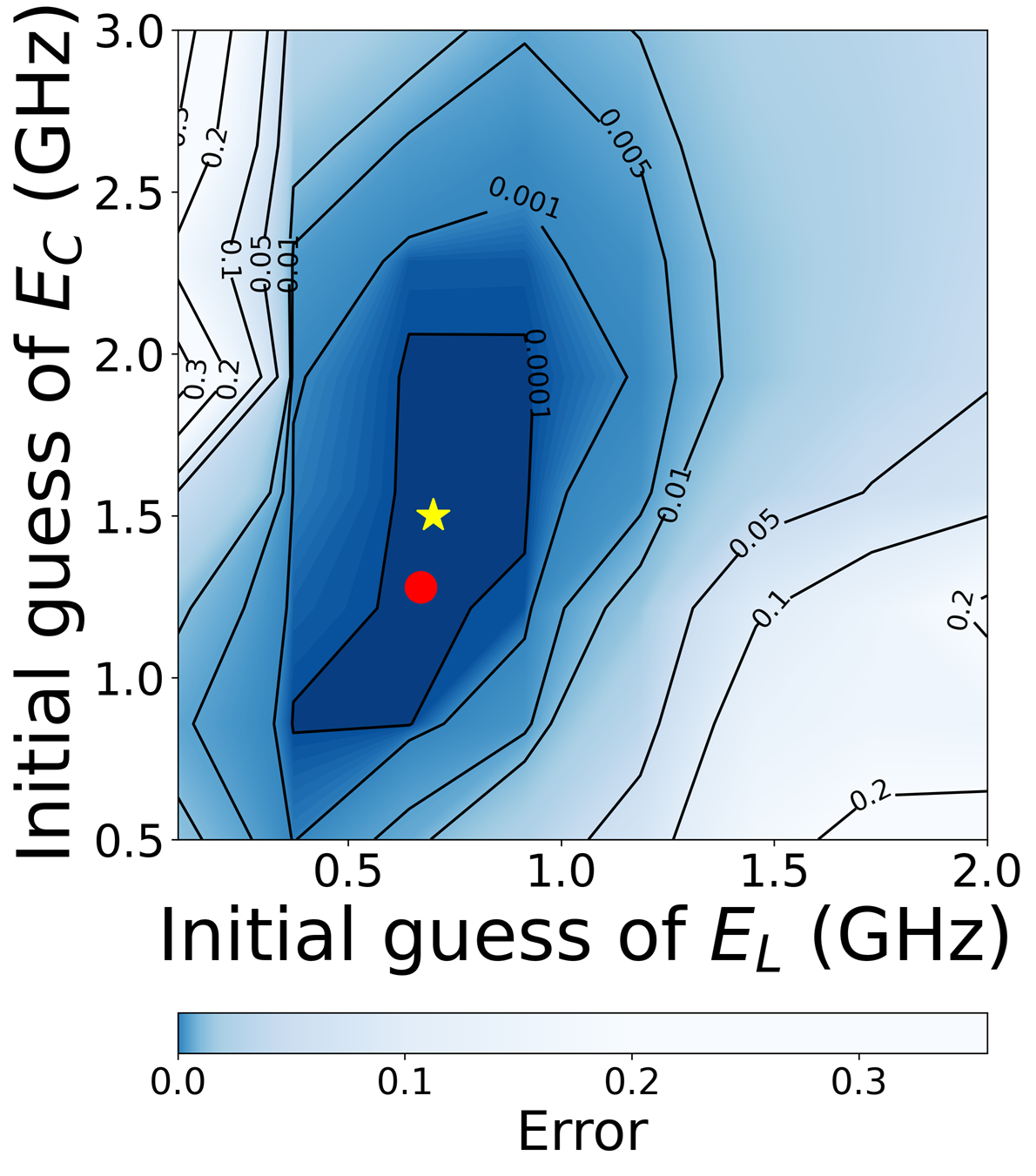} 
\caption{$Error$}
\label{fig:difference_ElEc}
\end{subfigure}
\begin{subfigure}{0.23\textwidth}
\includegraphics[scale=0.26]{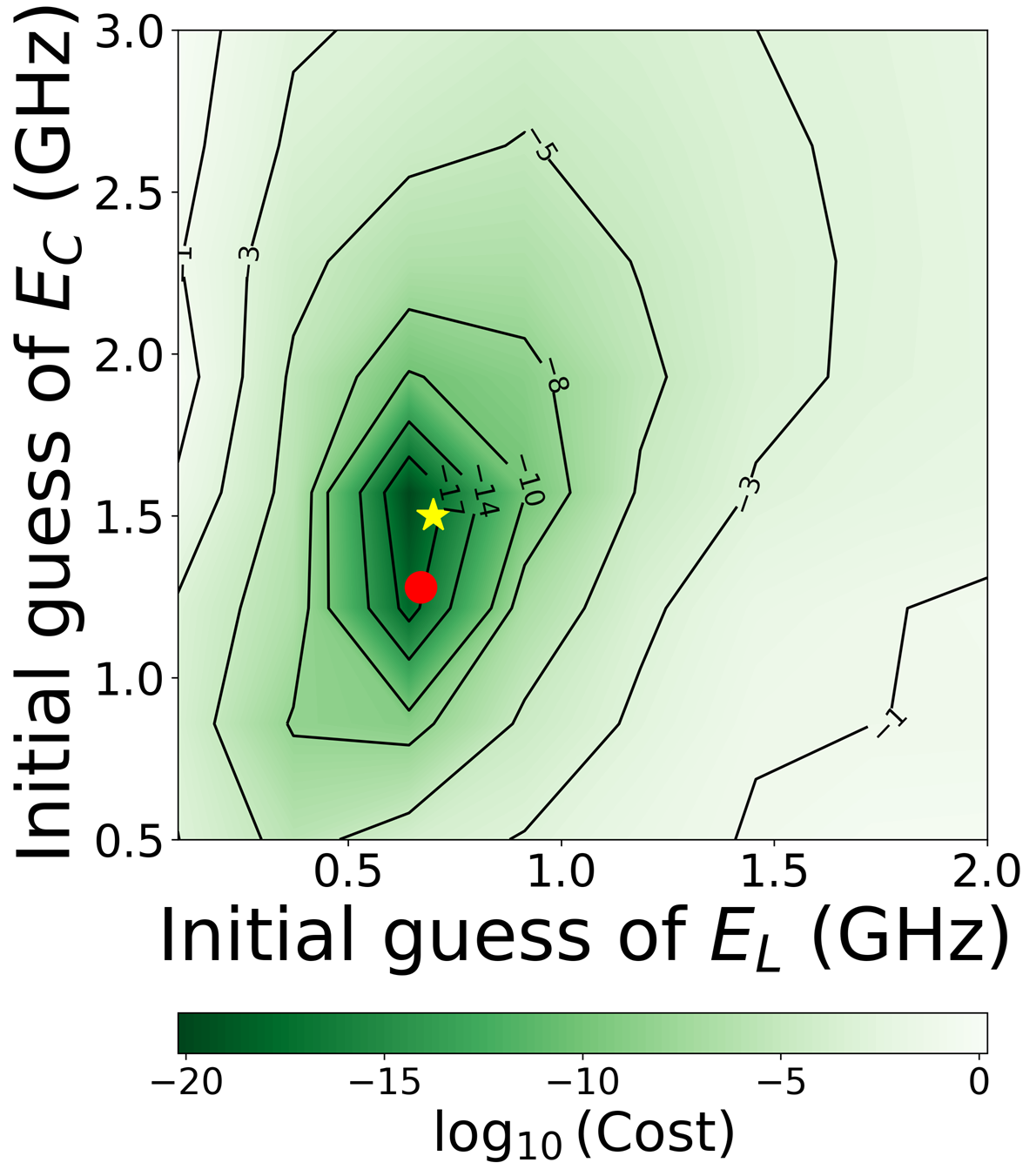}
\caption{$\log_{10}(Cost)$}
\label{fig:cost_ElEc}
\end{subfigure}
\captionsetup{format=plain, justification=RaggedRight}
\caption{
Same as Fig. \ref{fig:ElEj} but obtained by various values in the $E_L-E_C$ space with $E_J=7.05$ GHz.
The correct parameters is at $E_C=1.5$ GHz and $E_L=0.7$ GHz, as labeled by the yellow star symbol. The ML predicted results (red dots) is at $E_C=1.28$ GHz and $E_L=0.67$ GHz.}
\label{fig:ElEc}
\end{figure}

Figure \ref{fig:ElEj} and \ref{fig:ElEc} demonstrate the outcomes of utilizing arbitrary initial values to fit the fluxonium circuit parameters. For representational simplicity, the initial value of $E_C$ is fixed at 1.28 GHz in Fig. \ref{fig:ElEj} and the initial value of $E_J = 7.05$ GHz. in Fig.  \ref{fig:ElEc}. In Fig. \ref{fig:ElEj}, the horizontal and vertical axes represent the initial guess values of $E_L$ and $E_J$, respectively. The color density indicates the magnitude of $Error$ in panel (a) and $\log_{10}(Cost)$ in panel (b). Lower values of $Error$, represented by darker colors, indicate closer proximity to the true energy parameters. Similarly, lower $Cost$ values imply a reduced computational time for completing the fitting process. The truth qubits' parameters in this plot, denoted by yellow star symbol, is located at $E_L = 0.70$ GHz and $E_J = 6.50$ GHz. The resultant ML predicted qubits' parameters is $E_L=0.67$ GHz and $E_J=7.05$ GHz and plotted as the red dot. In Fig. \ref{fig:ElEc} presents an analogous plot with $E_C$ on the vertical axis and $E_L$ on the horizontal axis. The ground truth qubits' parameters in this plot, denoted by yellow dots, is located at $E_C = 1.50$ GHz and $E_L = 0.70$ GHz. The resultant ML predicted qubits' parameters is $E_C=1.28$ GHz and $E_L=0.67$ GHz and plotted as the red dot.  

A comparison between error and cost function reveals the critical importance of accurate prediction of the qubits' parameters $(E_C, E_J, E_L)$: even slight deviations from the correct values of qubits' parameters can result in a substantial increase in the $Cost$ metric. This sensitivity highlights the potential discrepancies between the predicted transition energies and the actual transition energies, which can significantly impact the fidelity of experimental control in quantum computational processes. For example, an error of 0.01 implies that the cost function can be as large 0.001, which results in a root mean square difference of 32 MHz in transition frequency. The difference is much larger than a typical linewidth of any qubit transitions. These findings emphasize the necessity for highly precise parameter estimation techniques in the development and operation of superconducting qubit systems.

\subsection{Comparison between ML Model and Traditional Method}\label{subsec:Comparison between ML Model and Traditional Method}

Analysis of contour plots, as shown in Figures~\ref{fig:ElEj}(b) and ~\ref{fig:ElEc}(b) reveals that the predictions generated by our ML model, represented by red dots, fall within the regions of lowest $Error$ and $Cost$, as indicated by the darkest areas of the plots. These ML-predicted values demonstrate close proximity to the true parameters with Error less than 0.0001 and $\log_{10}(Cost)$ less than -17 in this example. The strategic placement of these predictions suggests that utilizing them as initial values for traditional fitting processes would likely yield highly accurate final qubit parameters, even with a limited number of fitting iterations (e.g., five). This visualization provides a comprehensive comparison between conventional fitting methods and our ML approach, effectively illustrating the enhanced efficiency and accuracy of our initial parameter estimation technique across the multidimensional parameter space. The results underscore the potential of ML-augmented methods to significantly improve the characterization process for superconducting qubits, potentially reducing computational overhead and improving the precision of qubit parameter estimation in quantum computing applications.

To further quantify the comparative efficacy of our ML approach versus traditional fitting methods, we conducted a comprehensive analysis using 60 distinct parameter sets within the specified ranges of our test data. For each parameter $(E_C, E_J, E_L)$, we generated 512 uniformly distributed random initial values to seed the traditional fitting process. Following five fitting iterations, we computed the average $Error$ and $Cost$ metrics across these 512 initial value sets. These results were then juxtaposed with the fitting outcomes obtained using initial guesses provided by our ML algorithm. This systematic evaluation allows for a robust statistical comparison between the two approaches, elucidating the potential advantages of ML-driven initial parameter estimation in terms of both accuracy and computational efficiency.

\begin{figure*}
\includegraphics[scale=0.45]{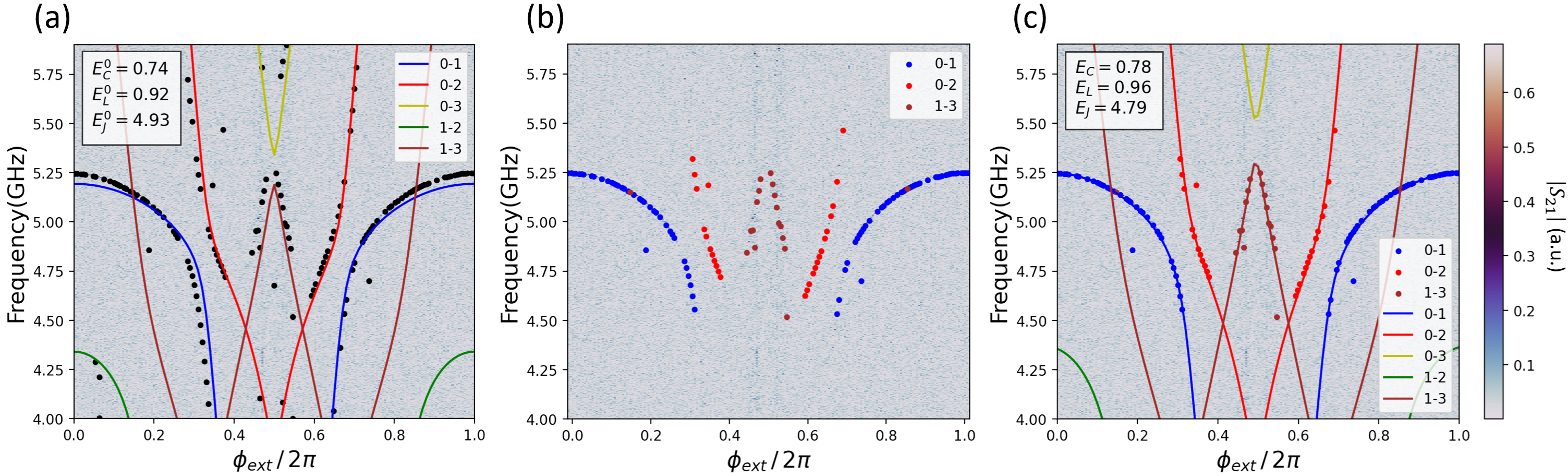}
\captionsetup{format=plain, justification=RaggedRight}
\caption{(a) The processed data (black dots) is input into the ML module, which generates predicted parameters as $E_C^0 = 0.74$ GHz, $E_L^0 = 0.92$ GHz, and $E_J^0 = 4.93$ GHz. The solid line represents the initial prediction by the ML model, and the different colored lines correspond to different state transitions. (b) The result of automating the identification of transitions within the processed data (for detailed methods, please refer to Sec.~\ref{subsec:An overview of our Method}). (c) The results obtained using this method with actual measurement data are $E_C = 0.78$ GHz, $E_L = 0.96$ GHz, and $E_J = 4.79$ GHz. Points of different colors indicate transitions between various energy levels. The solid lines represent the simulated spectrum based on the fitting results.
}
\label{fig:real_data}
\end{figure*}


Table \ref{table:1} presents a statistical summary of the fitting results across the 60 diverse parameter cases, including both the average values (AVG) and standard deviations (STD) for the $Error$ and $Cost$ metrics. The data demonstrate that the initial guesses provided by our ML model yield nearly one order of magnitude lower average values for both $Error$ and $Cost$ compared to those obtained using random initial values in the traditional fitting process. This marked improvement in both accuracy and computational efficiency underscores the efficacy of our ML approach in enhancing the parameter estimation process for fluxonium qubits. The reduced variability, as indicated by the smaller standard deviations, further substantiates the consistency and reliability of the ML-derived initial estimates. These results provide quantitative evidence for the potential of ML techniques to significantly improve the characterization of superconducting qubit systems, offering a more robust and efficient alternative to conventional methods.

\begin{table}[h!]
\centering
\renewcommand{\arraystretch}{1.2}
\setlength{\tabcolsep}{9pt} 
\begin{tabular}{ c  c  c  c } 
 \hline
    & & AVG & STD \\ 
 \hline\hline
 \multirow{2}{10em}{Random initial values} & $Error$ & 0.218 & 0.098 \\ 
    & $Cost$ & 0.146 & 0.130 \\
 \hline
 \multirow{2}{6.5em}{ML prediction} & $Error$ & 0.037 & 0.088 \\ 
    & $Cost$ & 0.024 & 0.083 \\
 \hline
\end{tabular}
\captionsetup{format=plain, justification=RaggedRight}
\caption{Comparison between the fitting results obtained by 512 random initial values after 5 iterations and results obtained by our ML method. The average values (AVG) of $Error$ and $Cost$ of the latter are nearly one order of magnitude smaller than the former.}
\label{table:1}
\end{table}

\subsection{The application on real fluxonium qubit energy spectrum}
\label{subsec:The application on real fluxonium qubit energy spectrum}

In this section, our ML approach is applied to actual measured data from a single fluxonium coupling to a 3D cavity. The measured spectrum first undergoes noise reduction and point selection before being analyzed with ML, as the black dots in Fig.~\ref{fig:real_data}(a). The initial guesses provided by the ML model are $E_C^0 = 0.74$ GHz, $E_L^0 = 0.92$ GHz, and $E_J^0 = 4.93$ GHz. To further analyze this data, we utilized the method described in Sec.~\ref{subsec:An overview of our Method} to determine the transition energy corresponding to each data point. Fig.~\ref{fig:real_data}(b) shows the classification results, where the blue, red, and brown dots represent the predicted transitions of 0-1, 0-2, and 1-3, respectively.

From Fig.~\ref{fig:real_data}(a), the initial predictions made by the ML model are shown to align closely with the actual data points. This provides us with a good starting point for successfully labeling the measured data belonging to various transition states and offers a reliable initial estimate for the subsequent fitting step. 

After the fitting process, the results obtained are shown in Fig.~\ref{fig:real_data}(c), which yields $E_C = 0.78$ GHz, $E_L = 0.96$ GHz, and $E_J = 4.79$ GHz. It is important to note that, apart from requiring user input for the positions of $\phi_{ext} = 0$ and $\phi_{ext} = \pi$, the rest of the method is fully automated. The initial guesses provided by the ML model are crucial, as they help reduce fitting time, improve accuracy, and serve as a key factor in automating the classification of each point into its corresponding transition.

It is worth mentioning that this method not only accelerates the characterization of qubit parameters but also significantly reduces the time required for measuring the spectrum. With only partial crucial information, this approach can effectively identify qubit parameters. Although the ML model is trained on a complete spectrum ranging from 4 to 8 GHz, Fig.~\ref{fig:real_data}(a) demonstrates that even when provided with information within the narrower range of 4 to 5.9 GHz, the model still generates a sufficiently accurate set of initial guesses to successfully fit the data.

To explore the limitations of this method, we doubled the current measurement interval, thereby reducing the number of measurement points. Additionally, we measured only half a period and then duplicated and symmetrized the graph. For example, in the measurement results shown in Fig.~\ref{fig:real_data}, when using only the data from the left half, symmetrizing it to the right half, and doubling the current measurement interval, the results were $E_C = 0.74$ GHz, $E_L = 0.98$ GHz, and $E_J = 4.98$ GHz. The differences compared to the parameters in Fig.~\ref{fig:real_data}(c) were less than 5\%. This illustrates that with a limited number of data points, it is still possible to accurately determine the qubit parameters, thereby significantly reducing the time required for spectrum measurement.

\section{Discussion}
\label{sec:Discussion}


Compared to the fitting method that employs random guessing of initial parameters, the results depicted in Fig.~\ref{fig:ElEj} and Fig.~\ref{fig:ElEc} indicate that the initial guess provided by the ML approach is primarily concentrated in the low fitting $Error$ and $Cost$ regime. This observation highlights the utility of the ML model, as it enables the estimation of appropriate initial parameters, thereby enhancing the efficiency of the fitting process, particularly in scenarios where new data are encountered and there is a lack of information on the expected initial parameters. The ability of the ML model to estimate appropriate initial parameters for the data fitting problem also creates an opportunity for automating the parameter characterization process during qubit experiments. Given that the qubit parameters can vary over time due to environmental factors such as oxidation, automating this process could significantly improve the efficiency of the routine.


Preparing training data for ML models can present several challenges. One of such challenges is determining the optimal number of data points required for each image to ensure accurate training. Insufficient data points can lead to inadequate model performance, while an excessive number of data points can result in overfitting, where the model becomes too closely tailored to the training data and fails to generalize well to new data. Additionally, the intensity of the spectrum in the data can be highly dependent on the initial state population, which can vary significantly across different experiments. Moreover, the number of excitations considered can also impact the information contained in the data, making it important to carefully control and account for these factors when preparing training data for ML models. 



It is worth noting that in many quantum computing systems, the parameter characterization of a single qubit is often sufficient, as each qubit typically has its own dedicated readout mechanism. This allows for independent calibration and operation of individual qubits, ensuring that their performance is optimized without requiring simultaneous characterization of multiple qubits. However, the characterization of multiple qubits, especially in systems where qubits are strongly coupled or share resources, presents a more complex challenge. Addressing this could be an important direction for future work, potentially enabling more efficient and scalable approaches for multi-qubit calibration in large-scale quantum computing architectures.

The proposed method offers flexibility and adaptability to different problems and experimental setups. It can serve as a general framework that can be applied to similar experiments with different datasets (for example, different kinds of qubits), where the same ML algorithm can be used with minor modifications to suit the specific data. This allows for the proposed method to be extended to more complicated problems. Furthermore, the flexibility of the proposed method enables researchers to incorporate additional features or parameters as needed, making it a versatile tool for a wide range of qubit experimental applications. Overall, the proposed method's flexibility and extendability make it a promising approach for ML-based tools to a diverse range of quantum engineering problems.


While the proposed method offers significant potential in facilitating accurate data fitting, it is important to acknowledge its limitations. One potential limitation of the proposed method is that if the data is excessively noisy or there is a lack of information, this could lead to poor prediction results. To mitigate this issue, one can tune the number of data points and observe the impact on the model's performance. It is worth noting that increasing the number of data points is not always a viable solution as it can lead to overfitting. Therefore, striking a balance between the number of data points and model complexity is essential to achieve accurate predictions. Despite these limitations, the proposed method still provides a promising foundation for developing more advanced ML-based approaches to qubit parameter characterization, with the potential to yield significant benefits in a variety of fields.

\section{Conclusion}
\label{sec:Conclusion}




In conclusion, we successfully introduced a machine learning-based method for the automatic characterization of fluxonium superconducting qubits, significantly enhancing the accuracy and efficiency of qubit parameter fitting in quantum computing systems. Compared to traditional manual parameter tuning, this method not only drastically reduces the required time but also improves fitting accuracy. By training a ML model based on the Swin Transformer V2 neural network and utilizing transfer learning, we achieved a high prediction accuracy of approximately 95.6\%, accurately estimating the initial values of the $E_C$, $E_L$, and $E_J$ parameters. This initial guess provides a robust starting point for subsequent labeling and fitting steps, effectively preventing prolonged fitting times or incorrect parameter estimation.

Moreover, we found that the method remains effective even when the amount of data required for measuring the spectrum is reduced, further optimizing time efficiency. This new method represents a significant breakthrough in the automatic characterization of fluxonium superconducting qubits and provides a promising framework that can be easily extended to characterize other types of qubits.

The source code is available after the manuscript is published at \href{https://github.com/kungsally/QCAI}{GitHub}, allowing researchers to reproduce our results and extend our work.

\acknowledgments{
We thank Prof. Chung-Yu Mou for the valuable discussions. DWW is supported by the National Science and Technology Council in Taiwan with the grant 110-2112-M-007-036-MY3. YHL is supported by the National Science and Technology Council in Taiwan with the grant 113-2119-M-007-008 and by the Ministration of Education Yuchan Young Scholar Fellowship. This work is supported by the National Center for Theoretical Sciences, the Higher Education Sprout Project funded by the National Science and Technology Council, and the Ministry of Education in Taiwan.
}

\nocite{*}
\bibliography{qcai}

\appendix

\end{document}